\documentclass[%
 reprint,%
 amssymb, amsmath,%
 aip,cha,%
]{revtex4-1}

\usepackage{docs}%
\usepackage{bm}%
\usepackage[colorlinks=true,linkcolor=blue]{hyperref}
\usepackage[d]{esvect}
\usepackage{graphicx}
\usepackage{dcolumn}
\usepackage{bm}
\usepackage{hyperref}
\usepackage[mathlines]{lineno}
\usepackage{verbatim} 
\usepackage{multirow} 
\usepackage{lipsum}
\usepackage{tabularx}%
\expandafter\ifx\csname package@font\endcsname\relax\else
 \expandafter\expandafter
 \expandafter\usepackage
 \expandafter\expandafter
 \expandafter{\csname package@font\endcsname}%
\fi
\begin{document}

\title{Thermo-remnant magnetization assisted switching response}%
\author{M. Tripathi $\textit{et al.}$ }
\affiliation{ UGC-DAE Consortium for Scientific Research, Indore-452001, India}

\begin{abstract}
	Here, we describe the prototype for precisely scaled temperature assisted magnetic recording scheme utilizing the characteristics of thermo-remnant magnetization (TRM) in bulk SmCrO$_3$ as a memory media. The TRM response can be exploited in completely reversible bipolar switching performances either by tuning the temperature only across a sharp thermal window $\Delta T_w$ = 10 K, or annealing isothermally with a single pulse of field $\ge$ 0.1 T by flipping the direction of magnetization from 0$^o$ to 180$^o$. The microscopic mechanism of TRM in SmCrO$_3$ is explored by means of diffuse scattering using high energy $\lambda = 0.4997\AA$ neutrons and thermal variation of coercivity in $T_N \pm \Delta T$ range. It is noticed that the huge heterogeneity reflected by the estimated average spin-spin correlation length of the fluctuations in microscopic spin configuration being $\le$ 90 $\AA$ and efficient domain wall pinning effect as the defect dimensions are approximately twice to the domain wall width, governs the peculiar characteristic of TRM in SmCrO$_3$.
\end{abstract}
\maketitle 

Thermo-remnant magnetization (TRM) is attributed to the virtue by which the material memorizes the imprints of previous magnetic field in which it was cooled across the occurrence of an order-disorder magnetic phase transition. Classically, TRM is related to the observation of permanent magnetization at room temperature in naturally occurring rockmagnets,\cite{neel1955some,chikazumi2009physics,spacey1958thermo,zhang2007study} therefore, from the palegeomic point of view, the TRM of rock magnets especially relative to the earth's magnetic field, are of particular interest. Recently, it has been understood that the remnant mangetization $\textit{memorized}$ across ordering temperature can be utilized in precisely scaled thermally assisted recording scheme \cite{thompson2000future,thiele2003ferh}. Contemporary magnetic recording mechanisms rely on total number of magnetically interacting grains per bit in particular grain size distributions for high density recording media\cite{weller2000high,thiele2003ferh}. The limitations in this process arise because for the grains with dimensions ($\nu$) less than critical limit, the energy to overcome the potential barrier $\nu H_{coercive}M_{saturation}$/2 between two energy minima corresponding to $\uparrow$ and $\downarrow$ magnetic states, can be of the order of thermal fluctuation energy k$_B$T, and therefore, intrinsic temperature fluctuations can spontaneously re-orient the magnetic moments. In TRM assisted recording mechanism, data is stored in a suitable recording media while cooling across transition temperature when the values of spontaneous domain magnetizaiton M$_s$ and magneto-crystalline anisotropy K$_U$ are sufficiently low to provide a highly stable magnetic data storage while avoiding the thermal fluctuation limit in the storage bit density $\rho_{bit}$/in$^2$ range\cite{thiele2003ferh}.

To meet the requirements of suitable memory media for thermally assisted recording, characteristics of significantly robust TRM across N\'eel temperature (T$_N$) in SmCrO$_3$ associated with ordering of Cr sub-lattices in canted antiferromagnetic  phase are studied here. The candidature can be justified with the fact the value of energy product (M$_{remnant}$ x H$_{coercive}$ ) significantly changes by factor of $\approx$ 1000 in a very sharp thermal window ($\Delta T_w$) of 10 K across T$_N$ and, as a consequence, fully reversible bi-stability is realized either by sweeping the temperature across $\Delta T_w$ or using magnetic field pulses $\ge$ 0.1 T. The high coercive fields in the vicinity of ordering temperature ensure the thermally stable scaling and consequently minimize the noise levels, whereas, low magneto-crystalline anisotropy and spontaneous magnetization ratio K$_U$/M$_S$ for temperature values close to transition, keeps the required magnitude of writing field sufficiently low\cite{guo2007uniform, zhang2000novel}. Evidently, understanding the origin and peculiar behavior of TRM in SmCrO$_3$ can serve as a recipe towards the goal of technologically feasible single phase TRM assisted recording base. The qualitative features of TRM are governed by the nature of defect states or any irregularity in infinite periodic crystal symmetry \cite{ozdemir2006magnetic,tung2006tunable}. Here, we have utilized the results of magnetometric measurements and high energy ($\lambda$ = 0.4997$\AA$) neutron diffraction in order to elucidate the mechanism presiding the origin of TRM in SmCrO$_3$ and to explore the microscopic factors responsible for the stability of TRM against thermal agitation. 

\begin{figure}[t!]
	\includegraphics[width=0.52\textwidth]{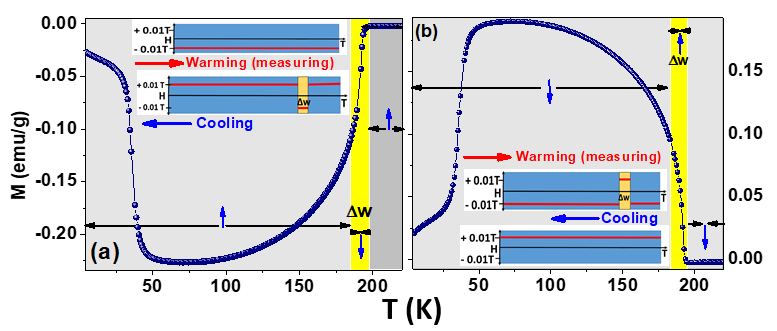}
	\caption{ Response of magnetic moment when system is cooled in non-uniform cooling field with respect to temperature, described as: cooling field $\mu_0H_{CF}$ is applied  across small temperature window in the vicinity of T$_N$, whereas, -$\mu_0H_{CF}$ is applied for rest of the temperature values. The measuring field ($\mu_0H_{MF}$ ) is also equal to -$\mu_0H_{CF}$ for all temperature values . The cooling and measuring scheme is: (a) $\mu_0H_{CF}|_{window} = - 0.0 1 T$ and $\mu_0H_{CF}|_{\forall T \neq T_{Window}} = + 0.0 1 T$ = $\mu_0H_{MF} = + 0.0 1 T$ (b) $\mu_0H_{CF}|_{window} = + 0.0 1 T$ and $\mu_0H_{CF}|_{\forall T \neq T_{Window}} = - 0.0 1 T$ = $\mu_0H_{MF} = - 0.0 1 T$ }
	\label{windowmt}
\end{figure}

\begin{figure}[t!]
	\includegraphics[width=0.49\textwidth]{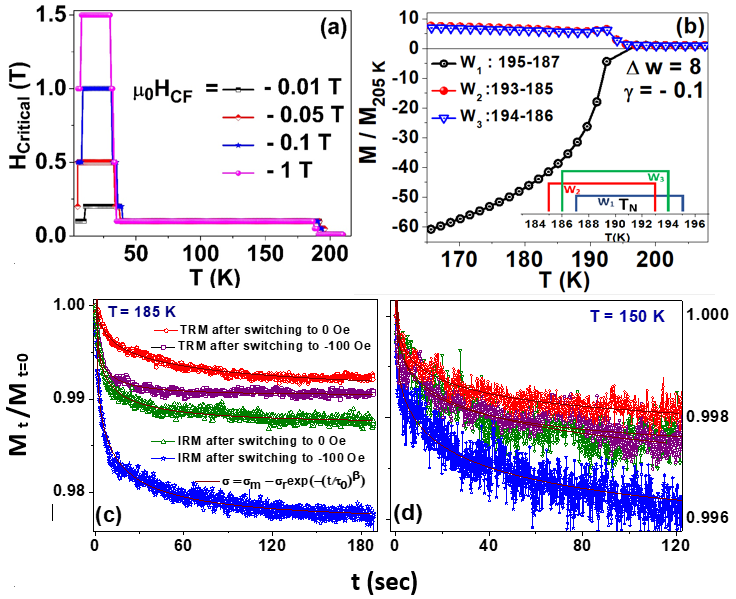}
	\caption{(a)Critical field as a function of cooling field $\mu_0H_{CF}$ and temperature. (b) The magnetization behavior in temperature window $\Delta W_T$ for constant width of window and $\gamma \equiv H_{CF}/H_{MF}$, but varying the range of temperatures for which cooling in presence of field H$_{CF}$ is performed. (c) and (d): Relaxation of thermo-remnant and iso-thermal remnant magnetization after switching field to zero and applying $\mu_0H = - 0.1$ T, a value equal and opposite to applied field used for initial magnetization of grains (see text), at T = 185 K and 150 K, respectively }
	\label{criticalnrelaxation}
\end{figure}

The sample preparation and description of magneto-crystalline structure are reported in ref\cite{tripathi2016phase, tripathi2017evolution} whereas, the details of experimental methodology with help of relevant reports \cite{rodriguez2001recent, kumar2010observation, tung2006tunable, tripathi2019role, fischer2002d4c, lynn1990resonance} are discussed in supplementary material (SM)\cite{SM}. The set of experiments presented in Fig.\ref{windowmt} illustrate how the value of cooling field across narrow temperature window ($\Delta T_w$) solely governs the sign of magnetization in whole temperature regime below T$_N$. We cooled the sample in presence of $\mu_0$H$_{CF}$= + 0.01 T for the entire temperature range from 300 K - 5 K except for a narrow temperature regime of $\Delta W_T$= 10 K across T$_N$= 191 K. In the temperature regime $\Delta W_T$= 10 K (195-185 K), we applied equal but opposite cooling field $\mu_0$H$_{CF}$= - 0.01 T. The magnetic moment was recorded during warming cycle in the presence of positive $\mu_0$H$_{MF}$= + 0.01 T. Clearly, the sign of moment is throughout negative following only the sign of cooling field which was present across a narrow window $\Delta T_w$ rather being bounded with the cooling field in rest of the temperature regime or on the measuring field(Fig.\ref{windowmt}(a)). The counterpart of this experiment is also performed as shown in Fig.\ref{windowmt}(b), in which the same scheme is followed but with reversed signs of cooling and warming fields as illustrated in schematics. These experiments reveal that major fraction of $\textit{memorized}$ magnetization is stored across the order-disorder transition only, which is called as thermo-remnant magnetization (TRM).
 Based on the series of further measurements, two additional key observations regarding to TRM in SmCrO$_3$ can be made: 

1. There exists a critical magnetic field H$_{cr}$$\equiv$H$_{cr}$(H$_{CF}$,T), defined as the minimum required  measuring (or warming) field to induce a magnetization with the same sign of applied field (H$_{cr}$.M$>$0). The variation of H$_{cr}$ as a function of temperature and cooling magnetic field  H$_{CF}$  is shown in Fig. \ref{criticalnrelaxation}(a). As the present magnetic state is a function of both cooling field H$_{CF}$ and measuring field H$_{MF}$, the critical field serves as an indicator of more dominating factor over sign of magnetization. When measuring field is less than  H$_{cr}$, the sign of magnetization follows the sign of cooling field irrespective of the value of measuring field. Note-worthily, the value of H$_{cr}$ for the temperature regime ranging from T$_{N}$ to spin reorientation phase transition (T$_{SRPT}$ = 34 K) is independent of cooling field, although it monotonically increases below T $\le$ T$_{SRPT}$. The plateau region in Fig.\ref{criticalnrelaxation}(a) shows that H$_{cr}$ is associated with the intrinsic magnetic structure and remains intact before the phase transition. The value of the H$_{cr}$ can be correlated with the average coercivity value of 851 (25) Oe in T$_N$-T$_N$-1 K temperature range, whereas, the coercivity at temperatures below T$_{N}$ ($\le 150 K$) are approximately an order higher than H$_{cr}$.\\
2. The magnetization behavior also depends on the temperatures incorporated in $\Delta W_T$ considering T$_N$ as mid reference temperature. It is noticed that $\Delta W_T$ is not an independent parameter rather it indirectly affects the functional dependency of M(T) on $\gamma \equiv H_{CF}/H_{MF} $. As illustrated in Fig.\ref{criticalnrelaxation}(b), for a fixed width $\Delta W_T= 8 K$, and for $\gamma$ = 0.1, three possible combinations of windows can be explored viz; w$_1$, w$_2$ and w$_3$. The magnetizion reversal is realized only when negative field is applied across  w$_1$ and, not in the case of w$_2$ and w$_3$. This peculiar observation reveals that even a slight change of 1 K in temperature values involved in $\Delta W_T$ regime can crucially alter the TRM . 
The magnetization of any single domain grain with volume $\nu$ has minimum energy configurations corresponding to two possible orientations $\theta = 0^o$ and 180$^o$, separated by potential barrier of height $\nu H_cM_s/2$. Depending on the value of $\nu $, the potential energy barrier can become comparable to the thermal fluctuation energy and moment direction can flip spontaneously where the change in magnetization can be characterized by relaxation time $\tau_0$. In the assemblage of such grains with remnant magnetization $\sigma_{r,t_0}$ at t$_0$, the time when field is set to zero (or switched to opposite direction), only the grains with $\tau \le \tau_0$ change their orientation. The decrease in remnant magnetization can be expressed as,
\begin{equation}
\sigma_{r,t} = \sigma_m - \sigma_{r,t_0}  exp({-t/\tau_0})^\beta 
\label{2}
\end{equation}
where, $\sigma_{m}$ represents the instantaneous ferromagnetic component and $\beta$ describes the distribution of potential energy barriers\cite{neel1955some, freitas2001magnetic, uehara2000relaxation}. The relaxation of thermo-magnetic remnant magnetization (TRM) and isothermal remnant magnetization (IRM) at T = 185 K and 150 K are exemplified in Fig.\ref{criticalnrelaxation}(c) and (d). To record the relaxation of TRM, we have cooled the sample from T$_N$ to the temperature at which measurement is performed, in presence of $\mu_0H = 0.01 $ T and data acquisition is started just after the magnetic field is switched to zero. Whereas, in case of IRM relaxation, no cooling field is applied, rather a magnetic field pulse $\mu_0H =  0.01$ T is applied at the measuring temperature for 100 seconds and data is recorded just after the pulse is removed. The same measurements are also recorded when instead of setting field value to zero, field is switched to $\mu_0H =  -0.01$ T, a value just opposite to the previously applied field. The relaxation curves are fitted with stretched exponential function described in equation.\ref{2} and estimated values of $\sigma_{r,t_0}$, $\beta$ and $\tau_0$ are given in SM. Clearly, TRM is significantly more stable against thermal fluctuations in comparison to the decay of IRM. In case of IRM, only the grains with H$_C$ less than 0.01 T are magnetized and consequently, the energy barriers between mangetized state and easy axis direction being the function of  H$_C$ and M$_s$ of a single grain, are also very small.  Whereas, most of the grains are magnetized during cooling across T$_N$ and the energy barrier is a function of  average coercivity of entire grain assemblage causing the observed larger relaxation times\cite{neel1955some}. In addition, switching the field in opposite direction instead of turning it to zero, prefers a comparatively slower decay of magnetization in both TRM and IRM processes. 

\begin{figure}[t!]
	\includegraphics[width=0.24\textwidth]{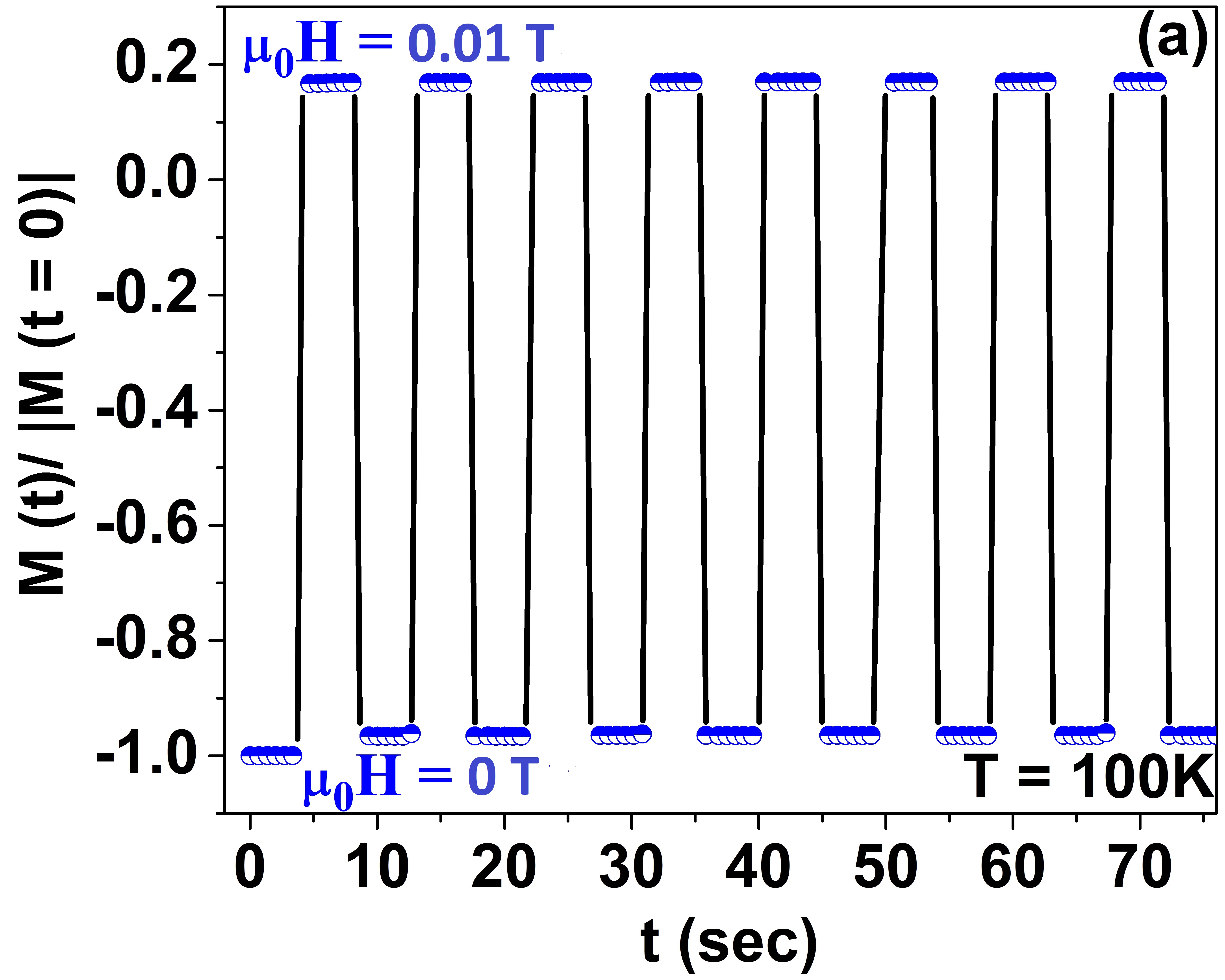}\includegraphics[width=0.25\textwidth]{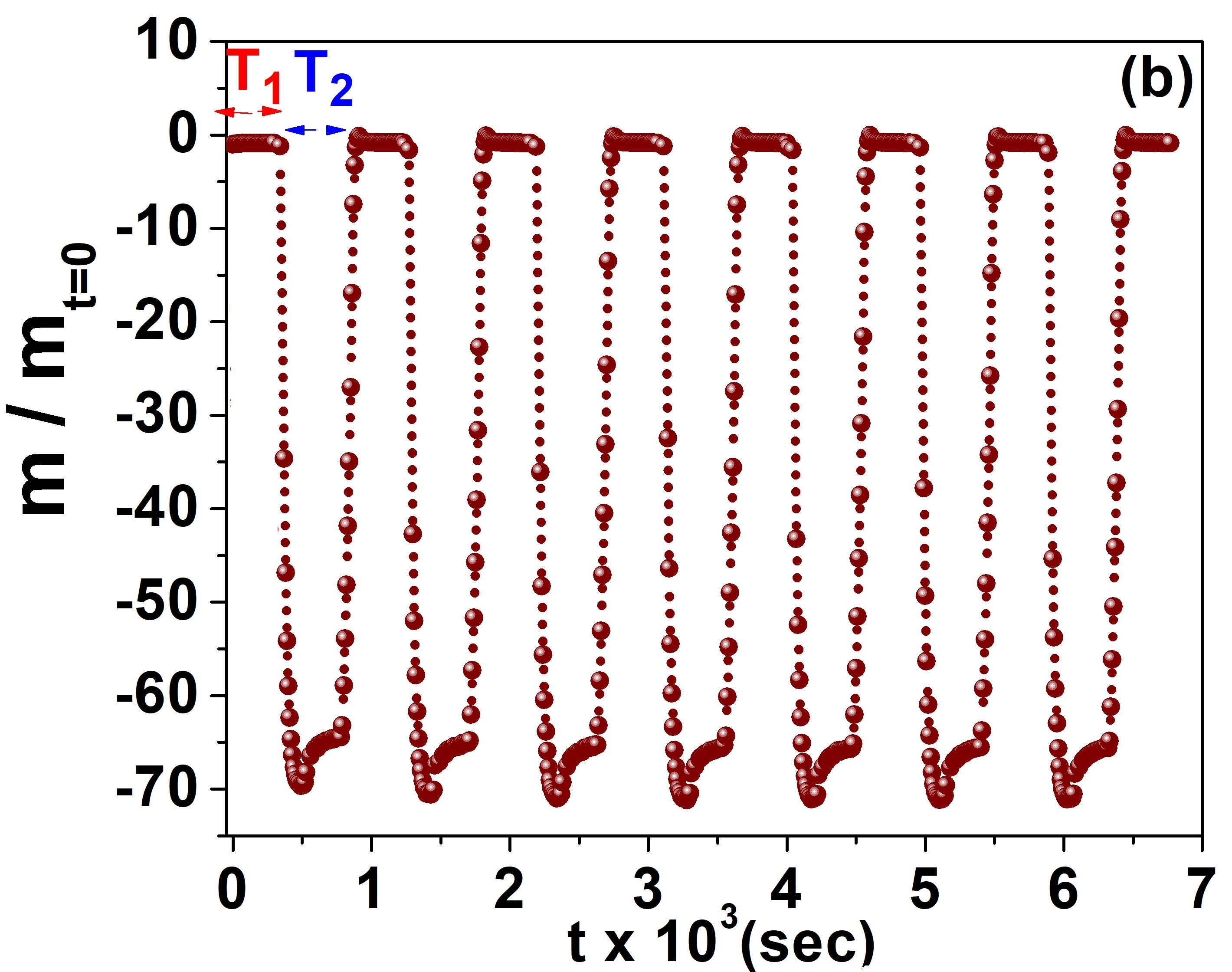}
	\caption{(a) Switching response with respect to H$_{app}$ and, (b) switching response with thermal variation across temperature window of $\Delta T_w$ = 10 K. (see text for details)}
	\label{HT}
\end{figure}

Owing to the stability of remnant magnetization for field values H$\le$H$_{cr}$ and its temperature and magnetic field dependent response, we can realize two types of bi-stability in this system. $\textbf{ (i) Switching response with respect to H$_{app}$}$ : In this case, the $\uparrow$ and $\downarrow$ magnetic moments serve as switching logic states. To record the switching response, we have cooled the sample in presence of - 0.1 T down to 100 K and using the magnetic field cycling 0 T - 0.1 T - 0 T - 0.1 T  ----- to change the sign of magnetic moment at a constant temperature T = 100 K recorded on time scale as shown in Fig\ref{HT}(a). The negligible switching delay and approximately full reversibility in magnitude indicate the possibility of its usage in data recording with two noticeable advantages over conventional methods: (i) only one magnetic field pulse is sufficient to perform bipolar switching action, as the $\downarrow$ magnetization state is realized without any application of field and, (ii) recorded information is volatile, the magnetic state returns to it's initially $\textit{memorized}$ state just after turning off the applied magnetic field with negligible time lag. $\textbf{ (ii) Switching response with thermal variation }$ : In second case, we have explored the reversal of magnetic moment polarity with thermal variation in narrow temperature window $\Delta T_w$ across T$_N$, as shown in Fig.\ref{HT}(b). The switching scheme for recording magnetization response in this case is as follows: sample is cooled in zero applied field down to 195 K (T$_1$) and then $\mu_0H_{CF}$ = -0.01T is applied and system is cooled down to 185 K (T$_2$) followed by recording the value of observed moment during warming cycle without changing the applied field value by sweeping the temperature in window T$_2$-T$_1$. In this switching mode, the switching process is governed exclusively by the change in temperature only, whereas the sole purpose of applied magnetic field is to assist in memorizing the previous logic state. The significant contrast in two stable logical states, full reversibility and very low magnetic field requirement can be utilized both in thermally assisted magnetic recording and automatic temperature sensors. 

The macroscopic interpretation of TRM in single domain assemblage is straight forward. The coercive forces of individual grains H$_C$ are proportional to spontaneous magnetic moment  M$_s$ and henceforth tends to zero when temperature approaches to Curie temperature. Therefore, in the vicinity of Curie temperature, very small value of applied magnetic field will be adequate to irreversibly magnetize the entire ensemble of grains. While cooling down the assemblage, the magnetizaiton will increase in proportion of $\textit{memorized}$ remnant magnetization. To understand the TRM in the interacting particle assemblage of multi-domain grains with a variable range of grain dimensions, 
\begin{figure}[t!]
	\includegraphics[width=0.49\textwidth]{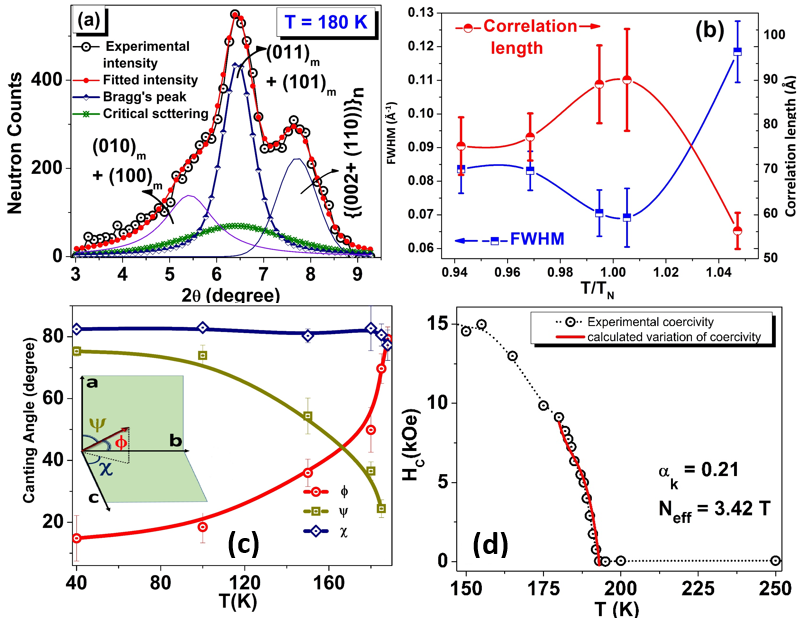}
	\caption{(a) De-convolution of magnetic (011)$_m$ + (101)$_m$ doublet feature into magnetic scattering profile and underneath broad critical scattering peak (b) Variation of calculated spin correlation length and FWHM with respect to temperature to reduced temperature T/T$_N$.(c) Variation of canting angle of magnetic moment with respect to three crystallographic axes. (d) Experimental and calculated variation of coercivity with respect to temperature.}
	\label{analysis}
\end{figure}
one needs to account many additional complex factors including grain-grain interaction, pinning and de-pinning effects at defect sites and volume anisotropy distribution corresponding to different sizes of clusters. 

 To estimate the magnetization fluctuations and spin correlation across T$_N$ we have utilized high energy ($\lambda = 0.4997~\AA$) diffuse neutron scattering. In the vicinity of critical point, the scattering intensity is strongly enhanced into forward direction at very small angles. The enhancement in scattered intensity is attributed to the sudden increment in amplitude of fluctuations of bulk magnetization being the order parameter of medium. In neutron diffraction profiles, just below ordering temperature, the observed peak profiles consist Bragg's Gaussian/Lorentzian feature superimposed over the diffused scattering profile\cite{bongaarts1975critical}. We have recorded the neutron diffraction patterns in the close neighborhood of T$_N$ to estimate the critical diffuse scattering. The doublet magnetic peaks (011)$_m$ + (101)$_m$ at 180 K is de-convoluted into two features:  Lorentzian magnetic scattering feature and, a broad critical scattering profile underneath, as shown in Fig.\ref{analysis}(a). The higher angle features corresponding to (002)$_n$ and (110)$_n$ Bragg's plane are fitted with Gaussian functions as they have only nuclear scattering contribution. The FWHM of nuclear or magnetic peaks are governed by temperature independent instrumental resolution function determined by measuring the neutron diffraction profile of standard sample. The FWHM of critical scattering profiles in $\AA^{-1} $ is estimated from similar deconvoluted total scattering profiles above and underneath at Bragg's (011)$_m$ + (101)$_m$ planes for temperature values in the close vicinity of ordering temperature T$_N$, as shown in Fig. \ref{analysis}(b). The correlation length is determined as follows :
$$ L(\AA) = 2\pi(FWHM(\AA^{-1}))^{-1}$$
Calculated values of spin correlation lengths are plotted as a function of reduced temperature T/T$_N$  in Fig.\ref{analysis}(b).  Average spin correlation length represents the length scale over which the fluctuations of spin configuration are correlated \cite{simons1997phase}. At temperature values higher than absolute zero, the correlation length is associated with the average dimension of local clusters which generally increases with thermal agitations up to transition temperature where sharp rise in fluctuations increases the correlation length significantly \cite{simons1997phase, witthauer2007phase}. In our system, the maximum spin-spin correlation length at N\'eel temperature is $\approx$ 90 $\AA$, revealing that the global order comprises an assemblage of clusters whose average dimension is less than 90 $\AA$. 

Now, we aim to investigate the role of defects comprising crystalline imperfections and magnetic irregularities on remnant magnetization. Considering the anisotropic defect as a quasi-harmonic perturbation for domain wall energy, Kronmuller et al.\cite{kronmuller1987theory, abraham1960linear}, expressed the coercive field as a function of temperature as:
\begin{equation}
H_c= 2K_u/{\mu_0J_s} \alpha_K -N_{eff}J_s 
\label{coercivity}
\end{equation}
where , K$_U$ is crystalline anisotropy constant, J$_s$ is saturation magnetization, $\alpha_K$ is a constant representing the nature of defects or pinning sites and N$_{eff}$ is an average parameter for dipolar interactions. The high field AFM behavior of SmCrO$_3$ is extrapolated by the straight line M = c$_0$ + m$\times$H$_{App}$ and subtracted to total magnetic moment in order to obtain the J$_s$.  We have estimated variation of canting angles over three crystallographic axes with respect to temperature using neutron powder diffraction. Thermal evolution of canting angles above T$_{SRPT}$ when the system orders in  $\Gamma$$_4$  (G$_x$,A$_y$, F$_z$) magnetic configuration is shown in Fig.\ref{analysis}(c). Clearly, the canting angle along easy axis (c$\parallel$z) $\psi$ is constant with respect to temperature and we can safely assume that magneto-crystalline anisotropy energy in $\Gamma$$_4$ phase is approximately a thermal constant. The value of magneto-crystalline anisotropy energy in G$_x$,A$_y$,F$_z$ configuration is $\approx $ 10$^6$ erg/mol\cite{cooke1974magnetic}. The calculated and experimental variation of coercivity across T$_N$ is shown in Fig.\ref{analysis}(d). The estimated values of $\alpha_K$ and N$_{eff}$ are = 0.21(7) and 3. 42 T(21), respectively. The high value of N$_{eff}$ reveals the strong dipolar exchange interactions in system. The value of $\alpha$ is correlated to the ratio of dimensions of defect ($\rho$) and domain wall width $\delta$. The situation $\rho$ = 2$\delta$ reveals high pinning efficiency and huge heterogeneity in SmCrO$_3$ \cite{givord1992coercivity}.

In the present work, the characteristics and underlying mechanisms of observed TRM in SmCrO$_3$ are studied, which provide the recipe for the so far un-accomplished single phase temperature assisted magnetic recording bulk media with meticulous storage density. Based on the functionality of TRM, we have proposed  two kinds of entirely reversible switching schemes; where the actuators for bipolar reversibility are magnetic field pulses and thermal variations, respectively. The microscopic insight of SmCrO$_3$ in the vicinity of N\'eel temperature is probed by means of diffuse neutron scattering and thermal variation of coercivity. Incorporation of the Kronm\"{u}lar's theoretical model to the thermal variation of coercivity reveals the large dimensions of defects and irregularity in periodic chains, which are approximately twice to the domain wall width. Using neutron diffuse scattering, the maximum average spin-spin correlation length ($\sim$ 90 $\AA$) at T$_N$ is observed. On the basis of these results, we argue that defects or grain boundaries are associated in breaking of continuous periodic symmetry within an average length of $\approx 90 \AA$ and, as a consequence efficient domain wall pinning effect and huge heterogeneity in system is generated which help in $\textit{memorizing}$ the information of previous magnetic state.  

We express our sincere thanks to Dr. H.  E. Fischer for help in the neutron diffraction experiment. Authors are grateful to Institut Laue-Langevin, Grenoble, France, for allowing the access of D4 diffractometer.

\bibliography{references}

\end{document}